\documentclass[twocolumn,twocolappendix]{aastex631}

\usepackage{graphicx}
\usepackage{amssymb}
\usepackage[fleqn]{amsmath}
\usepackage{times}
\usepackage{subfigure}
\usepackage[T1]{fontenc}
\usepackage{ae,aecompl}
\usepackage{textcomp}

\newcommand{\teff}{$T_{eff}$}

\begin{document}

\title{Discovery of the First Five Carbon-Enhanced Metal-Poor Stars in the LMC}

\author[0000-0001-7297-8508]{Madeline Lucey}
\affiliation{
Department of Physics \& Astronomy, University of Pennsylvania, 209 S 33rd St., Philadelphia, PA 19104, USA }

\author[0000-0002-0572-8012]{Vedant Chandra}
\affiliation{Center for Astrophysics $\mid$ Harvard \& Smithsonian, 60 Garden St, Cambridge, MA 02138, USA}

\author[0000-0002-4863-8842]{Alexander~P.~Ji}
\affiliation{Department of Astronomy \& Astrophysics, University of Chicago, 5640 S Ellis Avenue, Chicago, IL 60637, USA}
\affiliation{Kavli Institute for Cosmological Physics, University of Chicago, Chicago, IL 60637, USA}
\affiliation{NSF-Simons AI Institute for the Sky (SkAI), 172 E. Chestnut St., Chicago, IL 60611, USA}

\author[0000-0003-0174-0564]{Andrew R. Casey}
\affiliation{School of Physics \& Astronomy, Monash University, Wellington Road, Clayton, Victoria 3800, Australia}
\affiliation{Center for Computational Astrophysics, Flatiron Institute, 162 Fifth Avenue, New York, NY 10010, USA}

\author[0000-0002-1793-3689]{David Nidever}
\affiliation{Department of Physics, Montana State University, P.O. Box 173840, Bozeman, MT 59717, USA}

\author[0000-0002-6770-2627]{Sean Morrison}
\affiliation{Department of Astronomy, University of Illinois at Urbana-Champaign, Urbana, IL 61801, USA}

\author[0000-0003-3939-3297]{Robyn E. Sanderson}
\affiliation{Department of Physics \& Astronomy, University of Pennsylvania, 209 S 33rd St., Philadelphia, PA 19104, USA}

\author[0009-0000-9037-1697]{Slater Oden}
\affiliation{Department of Physics, Montana State University, P.O. Box 173840, Bozeman, MT 59717, USA}

\author[0000-0002-0900-9760]{Jos\'e G. Fern\'andez-Trincado}
\affiliation{Universidad Cat\'olica del Norte, N\'ucleo UCN en Arqueolog\'ia Gal\'actica - Inst. de Astronom\'ia, Av. Angamos 0610, Antofagasta, Chile}
\affiliation{Universidad Cat\'olica del Norte, Departamento de Ingenier\'ia de Sistemas y Computaci\'on, Av. Angamos 0610, Antofagasta, Chile}

\author[0000-0002-9269-8287]{Guilherme Limberg}
\affiliation{Department of Astronomy \& Astrophysics, University of Chicago, 5640 S Ellis Avenue, Chicago, IL 60637, USA}
\affiliation{Kavli Institute for Cosmological Physics, University of Chicago, Chicago, IL 60637, USA}

\begin{abstract}
A substantial fraction of metal-poor stars in the local Milky Way halo exhibit large overabundances of carbon. These stars, dubbed Carbon-Enhanced Metal-Poor (CEMP) stars, provide crucial constraints on the nature of the early universe including the earliest nucleosynthetic events. Whether these stars exist at similar rates in nearby galaxies is a major open question with implications for the environmental dependence of early chemical evolution. Here, we present the discovery of the first five CEMP stars in the Milky Way's largest dwarf companion, the LMC, using SDSS-V spectra from the BOSS instrument. We measure metallicities ranging from [Fe/H] = -2.1 to -3.2 and evolutionary state corrected carbon enhancements of [C/Fe] = +1.2 to +2.4, placing these stars among the most metal-poor and carbon-rich ever identified in the LMC. Their absolute carbon abundances and metallicities classify them as Group I CEMP stars, suggesting binary mass-transfer origins, though neutron-capture abundance measurements are required to confirm whether this classification scheme applies beyond the Milky Way. Although these stars were selected as the most promising CEMP candidates from the SDSS-V sample, likely biasing this initial sample toward higher absolute carbon abundances, their discovery suggests that previous null detections of CEMP stars in the LMC were caused by metallicity-sensitive photometric targeting biases against high [C/H] stars. A forthcoming analysis of the full spectroscopic sample will push to lower carbon abundances, providing a more complete census and enabling critical tests of whether environmental differences shape the formation channels and frequencies of CEMP stars in this system.

\end{abstract}



\section{Introduction} \label{sec:intro}

The chemistry of stellar atmospheres acts as a fossil record, as it generally remains unchanged over a star's lifetime except in cases of binary mass transfer or certain late-stage evolutionary processes in giant stars. Metal-poor stars formed from gas with few metals, likely before the universe had undergone significant metal enrichment. Their compositions therefore provide direct constraints on the nucleosynthetic yields of the first stellar generations and the chemical evolution of the early universe \citep{Beers2005,Frebel2015,Bonifacio2025}.

Early studies of metal-poor stars discovered that a significant fraction exhibit a large overabundance of carbon \citep[$\lbrack\rm{ C/Fe}\rbrack>0.7$;][]{Beers1992,Rossi1999,Beers2005,Lucatello2006,Christlieb2008}. The occurrence rate of these stars, referred to as carbon-enhanced metal-poor (CEMP) stars, increases at lower metallicities, making up 10-30\% of stars with [Fe/H]<-2 and $\approx$80\% with [Fe/H]<-4 in the local Galactic halo \citep{Lucatello2006,Lee2013,Placco2014,Yoon2018,Placco2018,Limberg2021}. 

CEMP stars are further divided into a number of sub-classes based on their neutron-capture element abundances \citep{Beers2005,Frebel2018}. Stars that exhibit enhancements in slow neutron-capture ($s$-process) elements (e.g., Ba) are dubbed CEMP-$s$ stars, typically defined as [Ba/Fe]>1. A small number of CEMP stars show enhancements in rapid neutron-capture ($r$-process) elements (e.g., Eu) which are called CEMP-$r$ stars. CEMP-$r/s$ stars exhibit enhancements in both $r$- and $s$-process elements \citep{Gull2018}, while CEMP-$i$ stars exhibit enhancements of intermediate neutron-capture ($i$-process) elements \citep{Frebel2018}. Lastly, CEMP-no stars do not exhibit any enhancements in neutron-capture elements, typically defined as [Ba/Fe]<0. CEMP-no and CEMP-$s$ are the most common subclasses while CEMP-$r$, CEMP-$r/s$ and CEMP-$i$ are more rare  \citep[e.g.,][]{Zepeda2023}.

The different sub-classes of CEMP stars are thought to be tied to different origin scenarios. CEMP-$s$ stars, which are more common at [Fe/H]>-2.5, are thought to be related to mass-transfer events from (post-)asymptotic giant branch (AGB) stars \citep{Lugaro2012,Placco2013}. Some of the strongest evidence for this hypothesis lies in the high binarity rate of CEMP-$s$ stars \citep{McClure1990,Preston2001,Lucatello2005,Bisterzo2010,Abate2015,Hansen2016,Jorissen2016}. \citet{Hansen2016}
find that CEMP-$s$ stars have a binarity rate as high as 82\%.

CEMP-no stars, on the other hand, have a lower binary fraction and are therefore thought to be unrelated to binary evolution \citep{Starkenburg2014,Hansen2016a,Arentsen2019}. Their excess carbon is thus likely a natal property of CEMP-no stars, indicating they formed from carbon-rich gas. Furthermore, given their high occurrence rate at the lowest metallicities, enhanced carbon yields may be characteristic of the first supernovae. Recent observations of carbon enrichment in early galaxies with $JWST$ provide further support for this scenario \citep{D'Eugenio2024}. Two theoretical explanations have been proposed for carbon overproduction in the first supernovae: rapid rotation in the first stars \citep{Chiappini2006,Meynet2006}, or explosion as faint supernovae \citep{Umeda2003,Nomoto2013,Tominaga2014}. Further studies of CEMP-no stars will help distinguish between these scenarios. 

One major clue may lie in the environmental dependence of CEMP stars. Contrary to naive expectations based on where the oldest stars are thought to concentrate \citep{El-Badry2018b}, CEMP stars are observed at lower rates in the inner Galaxy compared to the local halo \citep{Howes2014,Howes2015,Howes2016,Arentsen2021,Lucey2022}. Furthermore, while CEMP-no stars are found in ultra-faint dwarf galaxies at similar or higher rates to the Milky Way's halo, dwarf spheroidal galaxies exhibit a clear deficit \citep{Norris2010,Lai2011,Frebel2014,Skuladottir2015a,Salvadori2015,Ji2020,Lucchesi2024,Sestito2024,Chiti2025}. However, selection biases may drive these discrepant occurrence rates. Claimed deficits occur in more metal-rich environments where metallicity-sensitive photometric targeting methods are used to study the metal-poor stars, which are biased against selecting CEMP stars \citep{DaCosta2019,Chiti2020,Placco2025}. \citet{Lucey2023b} employed $Gaia$ low-resolution XP spectra to detect an unbiased, full-sky sample of candidate CEMP stars and found their density peaks towards the Galactic center. Furthermore, \citet{Lucey2023b} found a high density of candidate CEMP stars in the LMC and SMC, the Milky Way's largest dwarf galaxy companions. However, higher-resolution ($R\gtrsim$ 2,000) spectroscopic follow-up is required to confirm these findings. 

The LMC represents a unique opportunity to test the environmental dependence of CEMP. With a virial mass of $1-2\times 10^{11} M_{\odot}$ and stellar mass of $3\times 10^9 M_{\odot}$ \citep{Kim1998,vanderMarel2002,Erkal2019,Shipp2021,Watkins2024}, the LMC is only one order of magnitude less massive than the Milky Way. The metallicity distribution of the LMC peaks at [Fe/H]=-0.4 to -0.65, with only 10\% of stars having [Fe/H]<-1.0 \citep{cole2005,Nidever2020}. Detailed abundances for very metal-poor ([Fe/H]<-2) LMC stars have only recently been measured \citep{Nidever2020,Reggiani2021,Oh2024,Chiti2024,Ji2025,Limberg2025}, and no CEMP stars have yet been detected. However, confirming whether the LMC truly exhibits a lower occurrence rate of CEMP requires larger, unbiased samples.

The fifth-generation Sloan Digital Sky Survey \citep[SDSS-V;][]{Kollmeier2026} provides such a sample through its Magellanic Genesis program offering uniformly analyzed spectra for a large number of LMC stars, with limited targeting bias \citep{Nidever2026}. Here, we present the discovery of five CEMP stars in the LMC from this survey. In Section \ref{sec:Data}, we introduce the data including a short description of the target selection. Section \ref{sec:method} presents the spectroscopic analysis and in Section \ref{sec:discus}, we contextualize this discovery relative to LMC and Milky Way literature. Finally, Section \ref{sec:Sum} summarizes this work and the conclusions.

\section{SDSS-V BOSS Spectra}\label{sec:Data}

SDSS-V includes observations with the Baryons Oscillation Spectroscopic Survey (BOSS) spectrograph \citep{Smee2013}. These spectra have $R\approx 1800$ and wavelength coverage from 3650-9000 \AA. The spectra used in this work are from the 2.5m duPont telescope at Las Campanas Observatory (LCO) \citep{Bowen1973}. 
The spectra were reduced using \texttt{IDLspec2D v6\_2\_1} \citep[Morrison et al. in prep]{Bolton2012,Dawson2013} with coadds and continuum normalization performed by the \texttt{astra} pipeline (Casey et al. in prep). Overall, these spectra have similar resolution and wavelength coverage to numerous CEMP studies \citep[e.g.,][]{Yoon2018,Arentsen2021}, including previous iterations of SDSS with the SEGUE survey \citep{Lee2013}, of which the BOSS spectrograph is an upgrade with increased wavelength coverage and throughput. Furthermore, using SEGUE spectra, \citet{Lee2013} demonstrated the carbon abundance could be measured to within 0.35 dex down to SNR =15. The BOSS CEMP spectra presented in this work all have SNR>20. Therefore, these spectra have proven ability to detect CEMP stars. 

\subsection{Magellanic Genesis Program} \label{sec:lmcsel}

\begin{figure*}
    \centering
    \includegraphics[width=\linewidth]{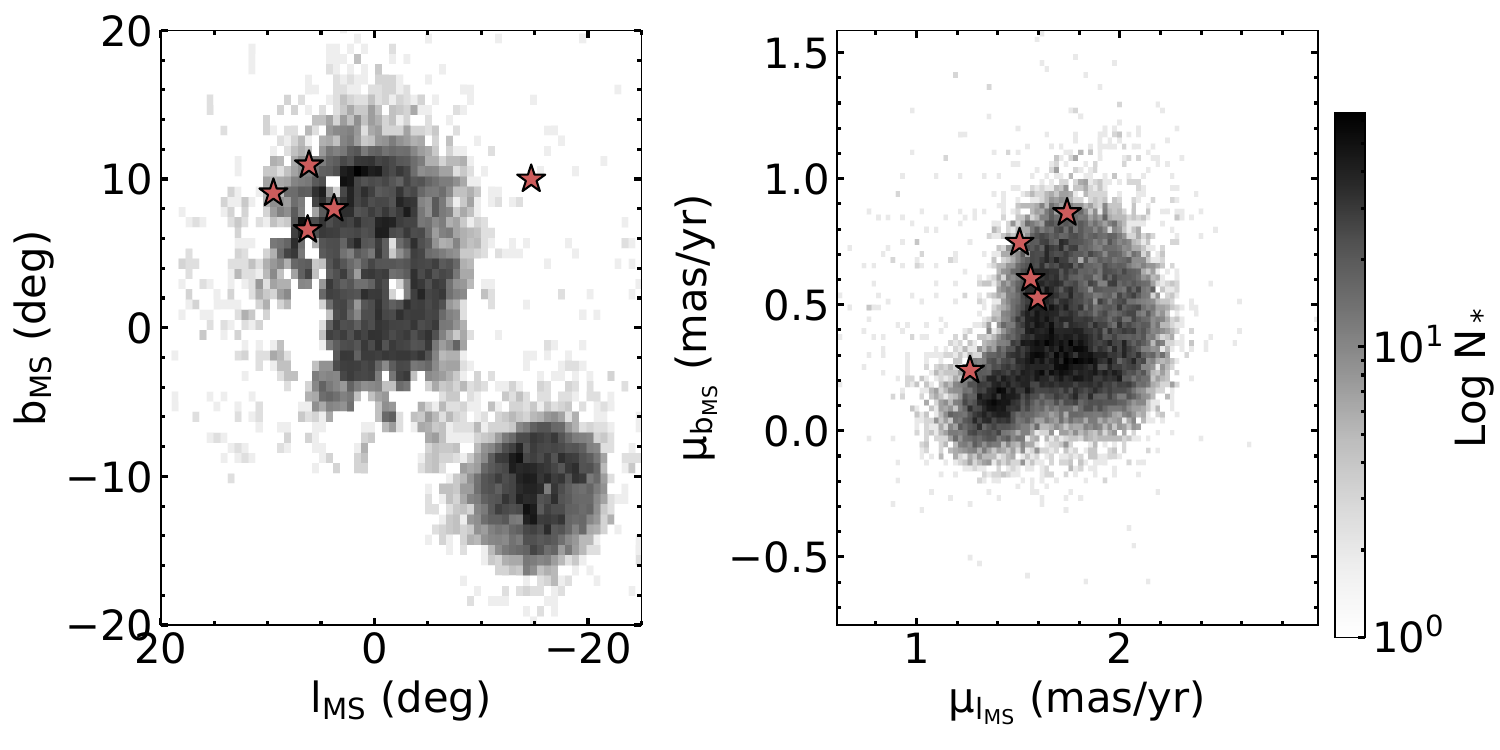}
    \caption{Distribution of sky positions (left panel), and proper motions (right panel) in Magellanic Stream coordinates \citep{Nidever2008} of the five CEMP stars (red stars) relative to the rest of the Magellanic Clouds sample in SDSS-V DR20. All five stars have sky coordinates and proper motions consistent with LMC membership.  }
    \label{fig:selec}
\end{figure*}

For a full description of the Magellanic Genesis program in SDSS-V, we refer the reader to \citet{Nidever2026}. Here, we briefly describe the targeting selection for the survey. The data used in this work specifically come from the red giant branch (RGB) selection which are selected using a 10-sided polygon in $Gaia$ G and BP$-$RP  with a bright limit of G=17.5 mag \citep[see Figure 7 in][]{Nidever2026}. The Magellanic Genesis program also applies a spatial selection, including only stars within 30$^{\circ}$ of (RA, DEC)=(80.8925$^{\circ}$, $-72.1849^{\circ}$). In addition, the program utilizes the following proper motion cut to remove contamination from Milky Way stars: 
\begin{equation}
    (\mu_{L_{MS}} -1.8)^2 + (\mu_{B_{MS}} - 0.4)^2 < 1.22
\end{equation}
where $\mu_{L_{MS}}$ and $\mu_{B_{MS}}$ are the proper motions in Magellanic Stream coordinates \citep{Nidever2008}.

Figure \ref{fig:selec} shows the sky positions and proper motions in Magellanic Stream coordinates \citep{Nidever2008} of our five CEMP stars (red stars) relative to the rest of the SDSS-V DR20 BOSS data. Four of our stars are located near the edges of the LMC disk, while the fifth star is more in the halo of the LMC. The proper motions are all consistent with being part of the Magellanic system. Radial velocities (RVs; see Table \ref{tab:table1}) as measured by the \texttt{BOSS-MINESweeper} pipeline \citep[see Section \ref{sec:MSparams};][]{Chandra2025} are also consistent with the LMC system. \texttt{BOSS-MINESweeper} RVs from LCO spectra are offset from those derived by the higher-resolution APOGEE spectra by an average of 9 km/s. This is due to a known wavelength calibration issue, for further discussion we refer the reader to Section 3.2 of \citet{Chandra2025}.

\section{Spectral Analysis} \label{sec:method}
In this work, we analyze five mid-resolution optical spectra from the SDSS-V BOSS instrument. First, we obtain estimates for the effective temperature (\teff), surface gravity (log $g$), and metallicity ([Fe/H]) using the \texttt{MINESweeper} pipeline, which produces stellar parameters for all stars in the SDSS-V halo survey \citep{Cargile2020,Chandra2025}. 
We then provide a secondary constraint on the metallicity by fitting the Ca~II infrared triplet. Last, we estimate the carbon abundances using spectral fitting to carbon molecular features. 

\subsection{\texttt{MINESweeper} Parameters} \label{sec:MSparams}

\begin{figure}
    \centering
    \includegraphics[width=\linewidth]{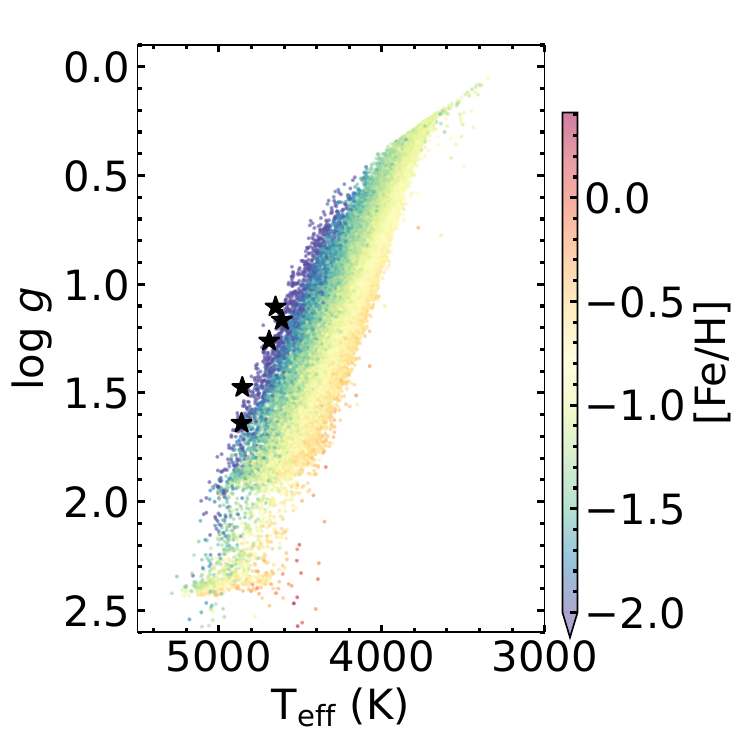}
    \caption{The \texttt{MINESweeper} stellar parameters for our five CEMP stars (black stars) relative to the rest of the SDSS-V DR20 Magellanic Genesis sample which is colored by metallicity. The five stars lay on the hotter edge of the red giant branch (RGB), consistent with having some of the lowest metallicities in the sample. }
    \label{fig:kiel}
\end{figure}

Stellar parameters for the SDSS-V halo survey are derived using the \texttt{MINESweeper} pipeline, producing the \texttt{BOSS-MINESweeper} value-added catalog\footnote{Public catalogs are available with each SDSS data release here: \url{https://www.sdss.org/dr19/data_access/value-added-catalogs/?vac_id=10007}} \citep{Cargile2020, Chandra2025}. For a full description of the \texttt{BOSS-MINESweeper} catalog, we refer the reader to \citet{Chandra2025}. As a short overview, the \texttt{MINESweeper} pipeline performs Bayesian inference to estimate the effective temperature (\teff), surface gravity (log $g$), [Fe/H], [$\rm{\alpha}$/Fe], projected rotation velocity ($\rm{V_{*}\, sin}\, \textit{i}$), initial mass ($\rm{M_{*,initial}}$), age, distance and extinction ($\rm{A_V}$). In addition to the BOSS spectra, the pipeline fits 2MASS, SDSS, Pan-STARRS, WISE, and $Gaia$ photometry, along with $Gaia$ parallaxes, where available \citep{Fukugita1996,Gunn1998,Skrutskie2006,Mainzer2014,Chambers2016,Gaia2021,Gaia2023}. The parameters are constrained to follow MIST isochrones, restricting solutions to physically-plausible regions of the parameter space \citep{Dotter2016,Choi2016}. The model spectra are synthesized in 1D LTE using the radiative transfer codes \texttt{ATLAS12} and \texttt{SYNTHE}  \citep{Kurucz1970,Kurucz1981,Kurucz1993}.

The nominal \texttt{BOSS-MINESweeper} analysis fits the spectra in the wavelength range 4750-5550 \AA.
Although this is a fraction of the available BOSS wavelength range, the model spectra have been empirically calibrated in this region to deliver reliable parameters. For this work,  we include a mask to remove the region between $5060$--$5180$~\AA{} where a $C_2$ Swan band dominates. The \texttt{MINESweeper} pipeline does not vary carbon and therefore cannot accurately model this region for carbon-enhanced stars. In order to ensure that this doesn't bias the results, we remove this region from the fit.

The \texttt{MINESweeper}-derived \teff\ and log $g$ of our five CEMP stars (black stars) relative to the rest of the SDSS-V DR20 Magellanic Genesis sample (colored by [Fe/H]) is shown in Figure \ref{fig:kiel}. The Magellanic Genesis sample is designed to target RGB stars and in order to achieve sufficient SNR, they tend to be luminous stars with low log $g$. Our five CEMP stars lie on the hottest end of the RGB, consistent with being very metal-poor.

\subsection{Sample Selection}

There are approximately 21,500 stars with reliable \texttt{MINESweeper} parameters in the SDSS-V DR20 Magellanic Genesis catalog.  Of these, 381 are CEMP candidates based on their \textit{Gaia} XP spectra \citep{Lucey2023b}. We then narrow it down to 5 stars for detailed analysis using preliminary fits to the CH G band and \texttt{MINESweeper} metallicities, picking the cleanest spectra with the lowest metallicities and largest [C/Fe]. There are certainly more CEMP stars within this sample that we do not analyze in this work. This paper presents the initial discovery of the first CEMP stars in the sample while the complete analysis of the sample, including estimates of the rate of CEMP enhancement and complete discussion of the selection function will be published in later work.

\subsection{Ca~II Triplet Metallicity Estimates } \label{sec:ca}

To validate and further constrain the metallicity estimates for these stars we fit the two strongest Ca~II infrared triplet lines at 8542 \AA\ and 8662 \AA. Throughout this work, we linearly interpolate synthetic spectra models. We used a synthetic spectrum grid of red giant stars computed in 1D LTE using Turbospectrum \citep{Plez2012}, spherical MARCS model atmospheres \citep{Gustafsson2008}, and the VALD atomic data \citep{Ryabchikova2015}. The grid was synthesized using 32,678 randomly chosen red giant surface gravities from $0 < \log g < 3.5$ with a large range of effective temperatures corresponding to a metal-poor Dartmouth isochrone \citep{Dotter2008}.

We utilize these models to estimate the [Ca/H]. See Appendix \ref{sec:app1} for more details on the spectral fitting. To convert this to [Fe/H], we assume a [$\alpha$/Fe] value. We do not use the [$\alpha$/Fe] from \texttt{MINESweeper} in order to achieve in independent estimate of the [Fe/H]. Instead, we use the median [Ca/Fe] value from \citet{Chiti2024} which measures abundances for stars in a similar metallicity range as this work (see Section \ref{sec:discus}) using high-resolution optical spectra from Magellan/MIKE. The median [Ca/Fe] abundance from that work is 0.24 dex with a standard deviation of 0.14 dex. The full range of [Ca/Fe] abundances is 0.02-0.48 dex. Therefore, we adopt [Ca/Fe] of 0.24 dex and estimate the [Fe/H] as [Ca/H]-0.24. 

The primary source of uncertainty for this [Fe/H] estimates comes from the [Ca/Fe] assumption. To be conservative we estimate the uncertainty on the [Ca/Fe] as 2 times the standard deviation of the [Ca/Fe] abundances from \citet{Chiti2024} which is 0.28 dex. We add this in quadrature with the variance from the spectral fitting. 

The [Fe/H] estimates from the Ca~II triplet generally agree with the [Fe/H] from \texttt{MINESweeper} with all of the differences being within 0.51 dex (see Appendix \ref{sec:app1}). For the final [Fe/H] estimates, we take the average of the two measurements. The uncertainty is then calculated using standard error propagation given two independent measurements.

\subsection{Carbon Abundance Estimates} \label{sec:c}

\begin{figure*}
    \centering
    \includegraphics[width=\linewidth]{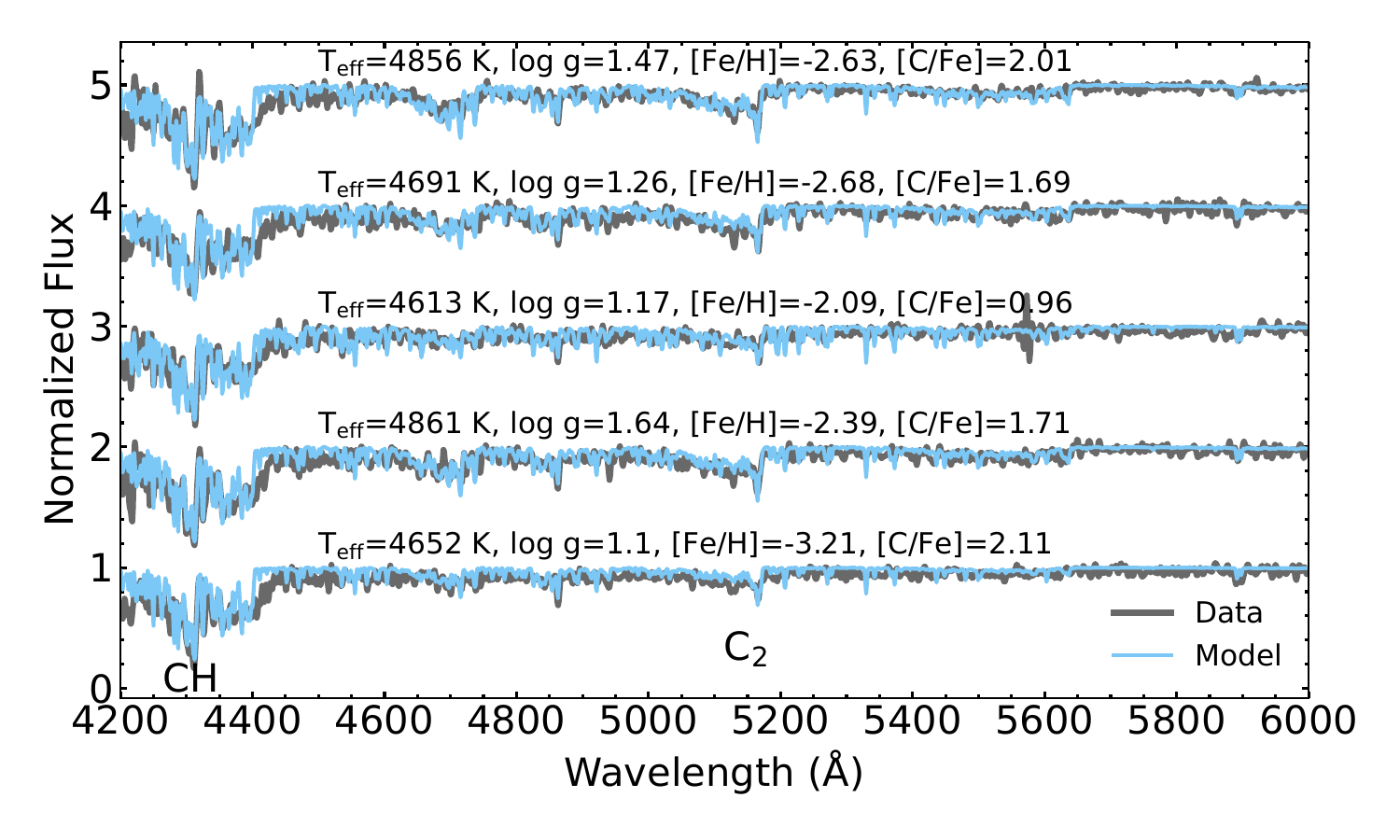}
    \caption{ A region of the BOSS spectra (grey) for the five CEMP stars with the best-fit synthetic spectral model (blue) overlaid. The spectral models shown are used for fitting the carbon abundance while the $\rm{T_{eff}}$ and log $g$ are fixed to the \texttt{MINESweeper} parameters. This region contains the CH G-band at 4305 \AA\ and the $\rm{C_2}$ Swan bands, with the strongest feature at 5165 \AA. The \texttt{MINESweeper} analysis also uses a subset of this region (4750-5550 \AA) for the stellar parameter analysis with a mask for the $\rm{C_2}$ band at $5060$--$5180$ \AA. }
    \label{fig:spec}
\end{figure*}

\begin{table*}
\caption{Sky Coordinates, and Stellar Parameters of the Five CEMP Stars }
\label{tab:table1}
\begin{tabular}{cccccccccc}
\hline\hline
SDSS ID & RA & DEC & SNR & RV & $\rm{T_{eff}}$ & log g & [Fe/H] & [C/Fe] & $\rm{[C/Fe]_{corr}}$  \\ 
& (deg) & (deg) & ($\rm{pixel^{-1}}$) & (km/s) & (K) & & & &\\
\hline
92278782 & 53.88 & -58.18 & 28.12 & $175^{+12}_{-12}$ & $4652^{+47}_{-42}$ & $1.10^{+.10}_{-.09}$ & $-3.21 \pm 0.21$ & $2.11\pm0.30$ & $2.41\pm0.30$\\
 96041179 & 96.65 & -65.10 & 20.93 & $244^{+17}_{-28}$ & $4861^{+71}_{-65}$ & $1.64^{+.17}_{-.15}$ & $-2.39\pm0.21$ & $1.71\pm 0.14$ & $1.85\pm 0.14$ \\
 98320880 & 90.49 & -64.07 & 25.01 & $291^{+11}_{-11}$ & $4614^{+53}_{-52}$ & $1.17^{+.12}_{-.12}$ & $-2.09\pm0.17$ & $0.96\pm 0.39$ & $1.23\pm 0.39$  \\
 98332219 & 102.15 & -61.90 & 20.46 & $317^{+10}_{-9}$ & $4692^{+110}_{-61}$ & $1.26^{+.20}_{-.13}$ & $-2.68\pm0.23$ & $1.69\pm 0.11$ & $2.05\pm 0.11$  \\
 98357416 & 94.57 & -60.84 & 31.74 & $295^{+8}_{-7}$ & $4856^{+52}_{-53}$ & $1.47^{+.10}_{-.11}$ & $-2.63\pm0.16$ & $2.01\pm 0.20$ & $2.17\pm 0.20$  \\
\hline
\end{tabular}
\tablecomments{The SDSS ID, right ascension (RA), declination (DEC), and Signal-to-Noise Ratio (SNR) of the BOSS spectra. The effective temperature ($\rm{T_{eff}}$), and surface gravity (log g) with associated uncertainties are from the \texttt{BOSS-MINESweeper} catalog. The derivations of the metallicity ([Fe/H], and  carbon abundance ([C/Fe]) are described in Section \ref{sec:ca} and  \ref{sec:c}, respectively. }

\end{table*}

Using the same synthetic spectra fitting procedure as above, we derive the carbon abundance in two regions, the CH G band between 4280-4350 \AA\ and a $\rm{C_2}$ Swan band between 5050-5200 \AA. We perform the fit to the regions simultaneously to achieve consistent estimates. Figures of the fits to each feature are included in Appendix \ref{sec:app2}. 

In Figure \ref{fig:spec}, we show the synthetic spectral fit (blue) overlaid on the observed spectra (grey) for each of the five CEMP stars. We label the CH G-band at 4505 \AA\ and the $\rm{C_2}$ Swan band at 5165\AA\ that we fit to derive the carbon abundance. The \teff\ and log $g$ are fixed to values from the \texttt{MINESweeper} pipeline. The [Fe/H] is the average between the \texttt{MINESweeper} value and the value estimated from the Ca~II infrared Triplet (see Section \ref{sec:ca}). In general, the model spectra fit the observed spectra well.

The largest source of uncertainty comes from uncertainty on the stellar parameters, \teff, log $g$ and [Fe/H]. To account for this, we recalculate the carbon abundance for each star given the stellar parameter uncertainties. We do this twice for each star, in the cooler star ($\rm{T_{eff,cool}=T_{eff}-\sigma_{T_{eff},l}}$, $\rm{log \textit{g}_{cool}=log \textit{g}-\sigma_{log \textit{g},l},}$  $[\rm{Fe/H]_{cool}= [Fe/H]-\sigma_{[Fe/H]}}$) and warmer star ($\rm{T_{eff,hot}=T_{eff}+\sigma_{T_{eff},h},}$  $\rm{log \textit{g}_{hot}=log \textit{g}+\sigma_{log \textit{g},h}}$, $\rm{[Fe/H]_{hot}= [Fe/H]+\sigma_{[Fe/H]}}$) scenario. We take the largest difference between the re-derived and fiducial values for each star as the respective stellar parameter propagated uncertainty. This is then combined in quadrature with the variance from the fits to the carbon molecular features (the square-root of the inverse Hessian of $\chi^2$) to calculate the total uncertainty.

As stars evolve up the RGB, the convective mixing leads to a depletion of carbon in the atmosphere. Therefore, we apply evolutionary state corrections to account for this effect on the derived abundances following \citep{Placco2014}. The correction is a function of log $g$, [Fe/H] and [C/Fe], with the correction being largest for stars with low log $g$, low [Fe/H] and low [C/Fe]. In general for our stars the corrections are on the order of 0.1-0.3 dex. Both the fit and corrected values are given in Table \ref{tab:table1}.

Four of our stars are robustly classified as CEMP with [C/Fe]>0.7 at greater than  3$\sigma$ significance. The other star (SDSS ID: 98320880) has only a 1$\sigma$ CEMP classification when including its evolutionary correction. With medium-resolution spectra, we are limited to >0.1 dex precision in carbon abundance, which makes robust CEMP classification difficult for stars with [C/Fe]$\lesssim$ 1 dex \citep{Arentsen2022}.

\section{Comparison to Literature} \label{sec:discus}

\begin{figure}
    \centering
    \includegraphics[width=\linewidth]{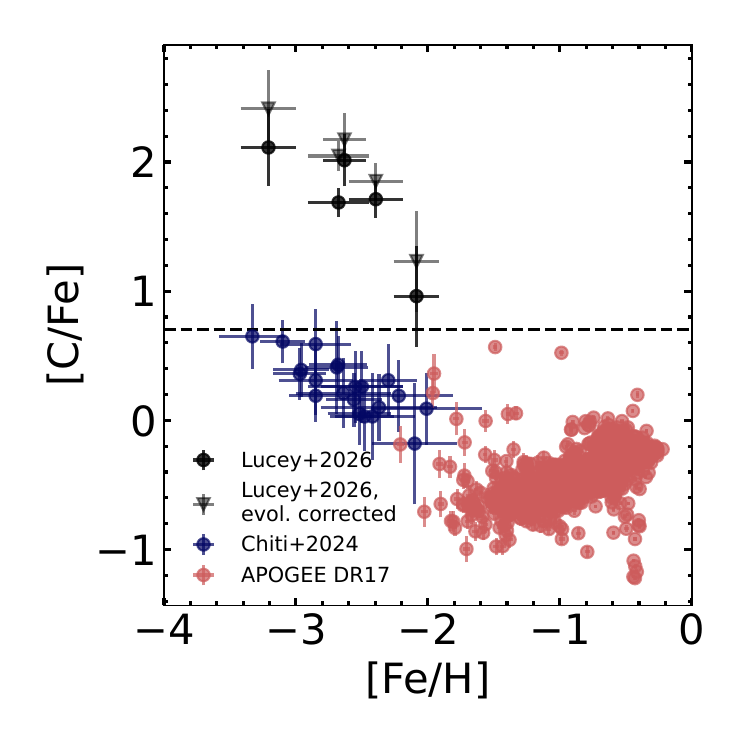}
    \caption{ Carbon abundances ([C/Fe]) as a function of [Fe/H] for the five CEMP stars compared to literature samples. Specifically, we compare to results from \citet{Chiti2024} which reports abundances for some of the most metal-poor stars known in the LMC, along with results from APOGEE spectra in SDSS-IV DR17. The commonly-used definition for CEMP ([C/Fe]>0.7) is shown as a black dashed line, with the five stars from this work being the only stars above said line.  }
    \label{fig:cfe}
\end{figure}

\begin{figure}
    \centering
    \includegraphics[width=\linewidth]{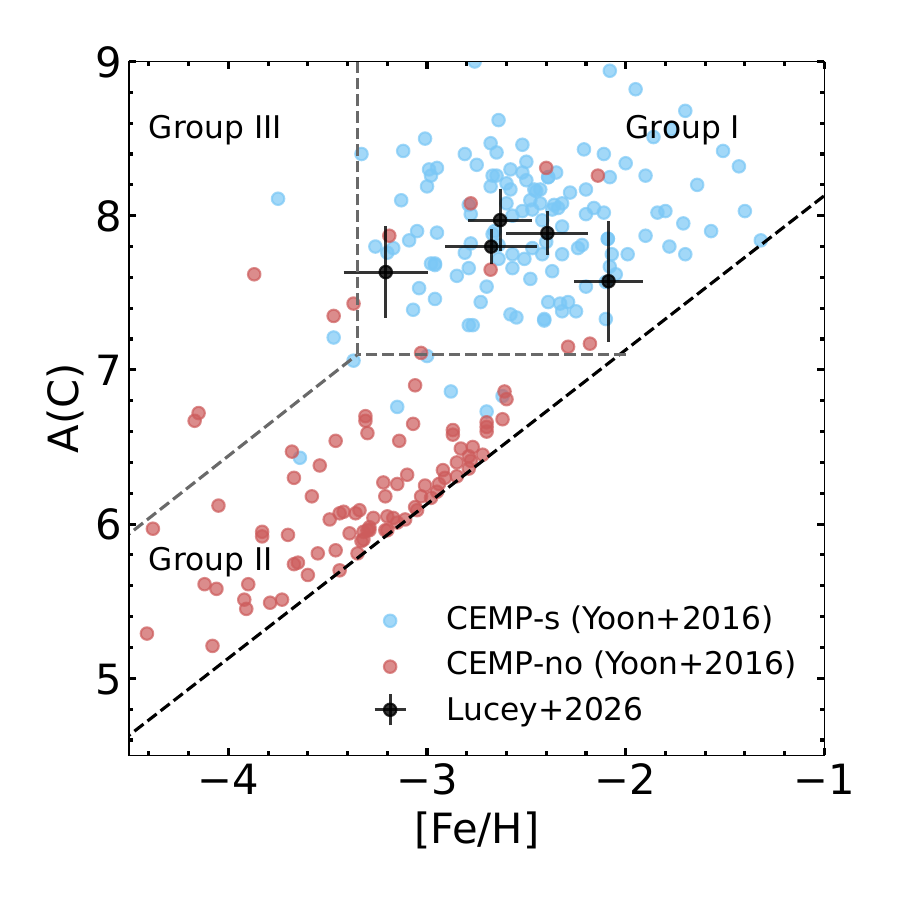}
    \caption{ The absolute Carbon abundance (A(C))  as a function of [Fe/H] for the five CEMP stars compared to a literature sample from the local Milky Way halo where the [Ba/Fe] abundances are also measured. The literature sample is seperated by [Ba/Fe] where stars with [Ba/Fe]>1 are labeled CEMP-s stars and are shown in blue whilte stars with [Ba/Fe]$\leq$ 0 are labeled CEMP-no and are shown in red. The black dashed line indicates [C/Fe]=0.7. The grey dashed lines indicate the classification system of CEMP introduced by \citet{Yoon2016}, where Group I is the top right section and are mainly CEMP-s, Group II and II are the bottom and top left sections and are mainly CEMP-no stars.    }
    \label{fig:yoon}
\end{figure}

To place our five CEMP LMC stars in context, we compare the measured abundances to literature samples from the LMC and the Milky Way. In Figure \ref{fig:cfe}, we show the [C/Fe] as a function of metallicity for our five stars (black circles) relative to previous works in the LMC. In addition, the [C/Fe] abundances that include the correction for evolutionary state  are shown in grey triangles. The dark blue points are  abundances from \citet{Chiti2024}, which targeted some of the most metal-poor stars in the LMC and also include the evolutionary state correction. In red, we show results from the SDSS-IV DR17 APOGEE catalog for the LMC \citep{Nidever2020,Abdurrouf2022}. The APOGEE abundances do not include an evolutionary state correction. It is possible that some of the APOGEE DR17 stars would be considered CEMP if the evolutionary state corrections were calculated and applied. However, the spectra would need to be visibly inspected to confirm the results as the APOGEE abundance pipeline is not designed to handle CEMP stars. 

From the literature and this work, we estimate the total number of stars with measured carbon abundances in the LMC with [Fe/H]<-2 to be on the order of around 40 \citet{Oh2024,Chiti2024,Nidever2020}. While previous works have found stars close to the CEMP limit when considering evolutionary corrections, these are the first robust detections. Previous works targeting very metal-poor stars in the LMC, use photometric selections which are known to be biased against detecting CEMP \citep{DaCosta2019,Chiti2020,Placco2025}. Our discovery confirms that the SDSS-V sample is more complete than the biased metallicity-sensitive photometric selections.  The five CEMP stars we report have substantially higher carbon abundances than ever previously measured in the LMC. 

As we are limited by the spectral resolution of our data, we do not measure neutron-capture element abundances of these stars. Therefore, we cannot assign these stars to specific CEMP subclasses. However, the absolute carbon abundance can be informative for classification \citep{Yoon2016}. In Figure \ref{fig:yoon}, we show the absolute carbon abundance (A(C)), which includes the evolutionary state correction, as a function of metallicity for our five stars compared to Milky Way CEMP stars with known neutron-capture abundances compiled in \citet{Yoon2016}.  The literature CEMP stars with enhanced $s$-process ([Ba/Fe]>1; CEMP-$s$ stars) are shown as light blue circles, while the CEMP-no stars with [Ba/Fe]<0 are shown in red circles. The black dashed line indicates [C/Fe]=0.7 dex. Individual uncertainties for the compiled abundances in \citet{Yoon2016} are not publicly reported but they report an average uncertainty of 0.25 dex.  The grey dashed lines indicate the different classes defined by \citet{Yoon2016}, with Group I in the top right, Group II in the bottom left and Group III in the top left. These groups are defined to capture the different structures in Figure \ref{fig:yoon}. Group I encapsulates the blob of CEMP-$s$ stars at higher metallicities and A(C). Group II represents the CEMP-no stars that have a roughly linear relationship between [Fe/H] and A(C), while Group III captures the CEMP-no stars with A(C)$\sim$7 at the lowest metallicities.

All five of our LMC CEMP stars lie in the Group I region which indicates they may be CEMP-$s$ stars. The preference for Group I stars in this sample reflects our selection of the most promising CEMP candidates with the strongest carbon features and should not be interpreted as indicative of the intrinsic CEMP population ratios in the LMC.  Furthermore, it remains unclear whether this classification system is directly applicable beyond the Milky Way given differences in chemical evolution across galaxies of varying mass, and there is already discussion of environment-dependent CEMP classifications in which the [C/Fe] threshold scales with the mean [C/Fe] of the host system \citep[e.g.,][]{Sestito2024}. Higher-resolution spectra to confirm neutron-capture element abundances will be essential for testing whether and how these classification schemes apply in the LMC.

The Group I classification of all five stars tentatively points toward AGB binary mass transfer as the dominant enrichment channel, though higher-resolution data are required to confirm their neutron-capture abundances before faint supernovae \citep{Tominaga2014}, or  spinstars \citep{Meynet2006} can be ruled out. Future work performing a comprehensive analysis of the full sample and measuring the CEMP occurrence rate in the LMC, accounting for selection effects, will provide deeper insight into early star formation in this system. Notably, our spectroscopic survey already recovers CEMP stars that metallicity-sensitive photometric surveys systematically miss, and future work extending to lower carbon abundances will provide a truly complete census, allowing us to confirm or rule out the existence of CEMP-no stars in the LMC far more definitively than metallicity-sensitive photometric targeting methods have achieved.

\section{Summary and Conclusions}\label{sec:Sum}

CEMP stars provide crucial constraints on early universe nucleosynthesis \citep{Chiappini2006,Tominaga2014,D'Eugenio2024}. Whether the occurrence rate of CEMP stars at low metallicities is dependent on the environment is an ongoing mystery that has interesting implications for the nature of the early universe and star formation. While CEMP stars are abundant at low-metallicities in the local Milky Way halo, and ultra-faint dwarf galaxies, there is an observational deficit in dwarf spheroidal galaxies, including no previous detections in the LMC \citep{Lucchesi2024,Sestito2024,Chiti2024,Ji2025,Limberg2025}. However, it is unclear whether this is a true discrepancy or whether photometric metallicity selection introduces biases \citep{DaCosta2019,Chiti2020,Placco2025}.  

In this work, we use data from the SDSS-V Magellanic Genesis program to search for CEMP stars in the LMC. We identify CEMP candidates within the data using the catalog from \citet{Lucey2023b}. Of those for which reliable \texttt{MINESweeper} parameters were derived, we select five stars with the strongest carbon molecular features to derive abundances. Specifically, we measure carbon using synthetic spectra fits to the CH G band and a $\rm{C_2}$ Swan band. 

With this method, we discover the first five CEMP stars in the LMC. Based on their absolute carbon abundances and metallicities, these stars are likely CEMP-$s$ stars given the classification presented in \citet{Yoon2016}. However, it is unclear whether this classification would apply beyond the Milky Way and therefore further observations to confirm the neutron-capture elements are required. 

The discovery of these five stars represents a crucial first step towards understanding the chemistry of the most metal-poor stars in the LMC. Their detection, where previous very metal-poor star searches found none \citep{Oh2024,Chiti2024}, provides further evidence that metallicity-sensitive photometric targeting systematically misses CEMP stars due to its bias against high [C/H] stars. Future work, extending our analysis to the entire SDSS-V LMC sample, including lower carbon abundances, will provide a more complete census than photometric metallicity targeting can achieve, enabling us to definitively measure the CEMP occurrence rate and confirm or rule out the existence of CEMP-no stars in the LMC.  In total, this, along with other upcoming LMC surveys \citep[e.g., 4MOST 1001MC;][]{Cioni2019}, will bring new understanding to nucleosynthesis in the early universe, especially with respect to environmental effects. 

\software{Astropy \citep{astropy:2013,astropy:2018},
Matplotlib \citep{matplotlib},
IPython \citep{ipython},
Numpy \citep{numpy}, 
Scipy \citep{scipy},
Gala \citep{gala,gala2}
}

\section*{Acknowledgements}

\begin{acknowledgments} 
This material is based upon work supported by the National Science Foundation under Award No. 2303831. 

Funding for the Sloan Digital Sky Survey V has been provided by the Alfred P. Sloan Foundation, the Heising-Simons Foundation, the National Science Foundation, and the Participating Institutions. SDSS acknowledges support and resources from the Center for High-Performance Computing at the University of Utah. SDSS telescopes are located at Apache Point Observatory, funded by the Astrophysical Research Consortium and operated by New Mexico State University, and at Las Campanas Observatory, operated by the Carnegie Institution for Science. The SDSS web site is \url{www.sdss.org}.

SDSS is managed by the Astrophysical Research Consortium for the Participating Institutions of the SDSS Collaboration, including the Carnegie Institution for Science, Chilean National Time Allocation Committee (CNTAC) ratified researchers, Caltech, the Gotham Participation Group, Harvard University, Heidelberg University, The Flatiron Institute, The Johns Hopkins University, L'Ecole polytechnique f\'{e}d\'{e}rale de Lausanne (EPFL), Leibniz-Institut f\"{u}r Astrophysik Potsdam (AIP), Max-Planck-Institut f\"{u}r Astronomie (MPIA Heidelberg), Max-Planck-Institut f\"{u}r Extraterrestrische Physik (MPE), Nanjing University, National Astronomical Observatories of China (NAOC), New Mexico State University, The Ohio State University, Pennsylvania State University, Smithsonian Astrophysical Observatory, Space Telescope Science Institute (STScI), the Stellar Astrophysics Participation Group, Universidad Nacional Aut\'{o}noma de M\'{e}xico, University of Arizona, University of Colorado Boulder, University of Illinois at Urbana-Champaign, University of Toronto, University of Utah, University of Virginia, Yale University, and Yunnan University.  

This work has made use of data from the European Space Agency (ESA) mission
{\it Gaia} (\url{https://www.cosmos.esa.int/gaia}), processed by the {\it Gaia}
Data Processing and Analysis Consortium (DPAC,
\url{https://www.cosmos.esa.int/web/gaia/dpac/consortium}). Funding for the DPAC
has been provided by national institutions, in particular the institutions
participating in the {\it Gaia} Multilateral Agreement.

\end{acknowledgments}

\appendix
\section{Ca~II infrared Triplet Metallicities} \label{sec:app1}

\begin{figure*}
    \centering
    \includegraphics[width=\linewidth]{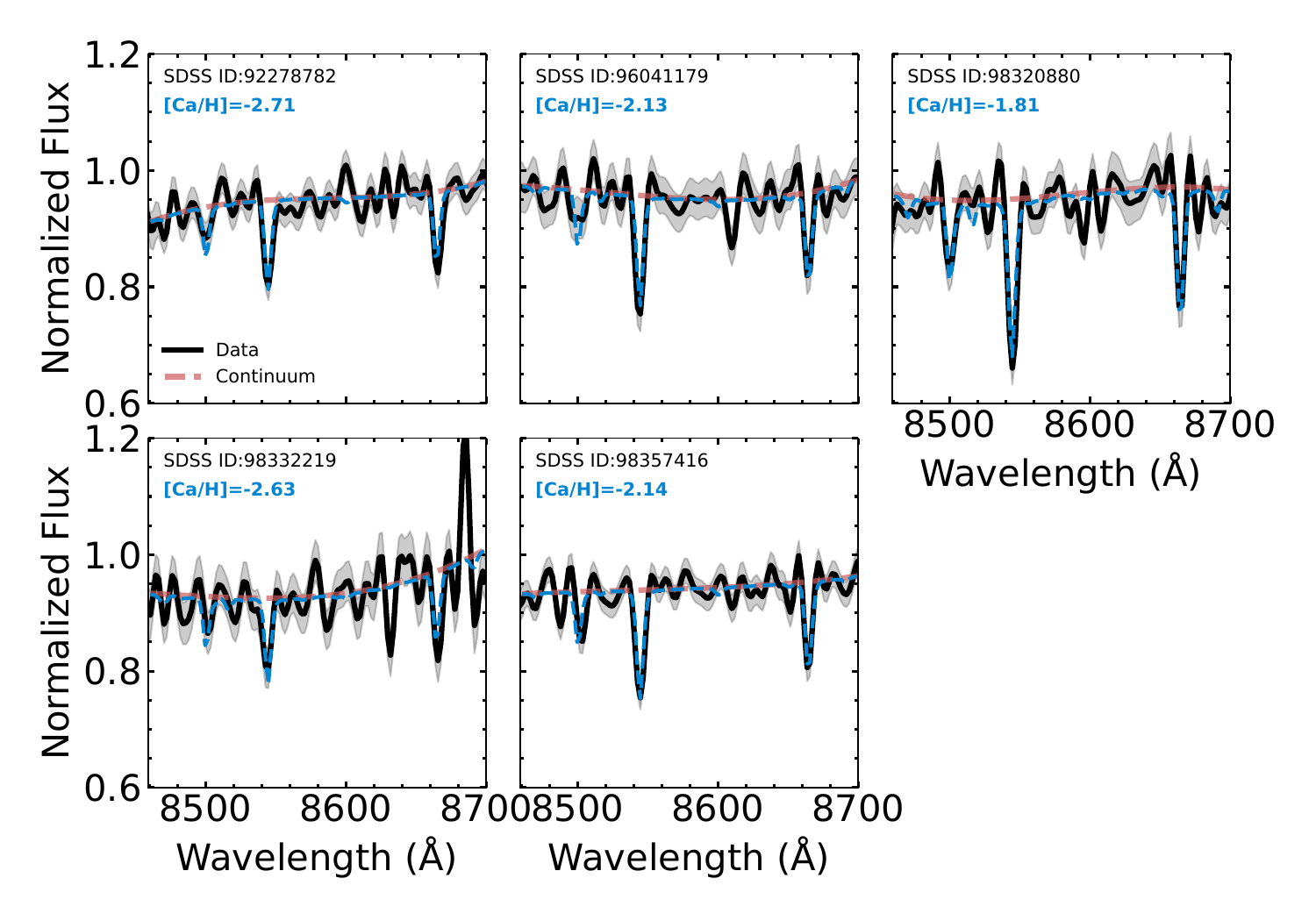}
    \caption{The fits to the Ca~II infrared triplet for the five CEMP stars. The black line with the grey shaded region give the observed normalized flux and the corresponding uncertainty. The blue dashed lines is the best-fit synthetic spectra. As the fit is highly sensitive to the continuum normalization, we fit the model continuum to best match the observed spectra. The red dashed line gives the continuum normalization for the best-fit  model.  }
    \label{fig:catrip}
\end{figure*}

\begin{figure}
    \centering
    \includegraphics[width=\linewidth]{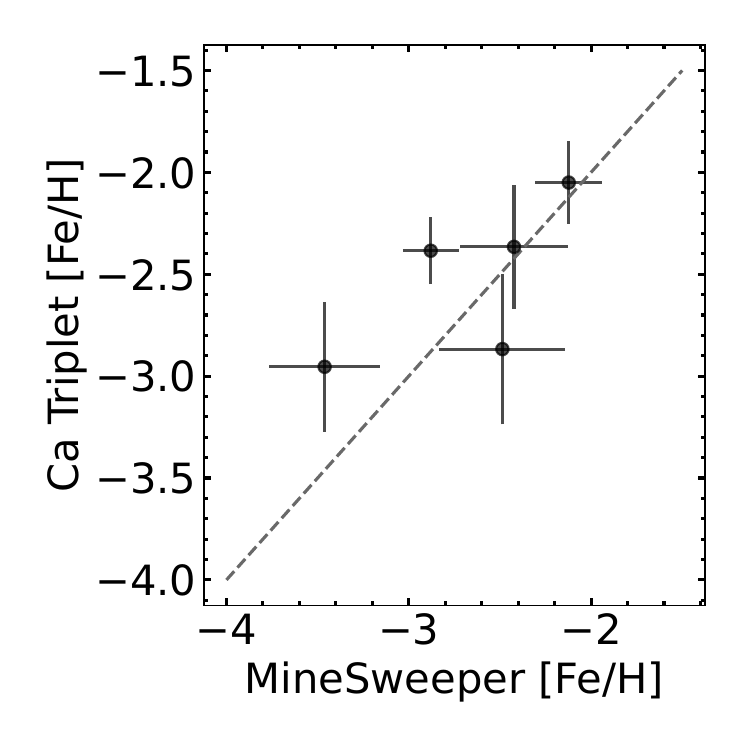}
    \caption{The metallicities derived from the Ca~II triplet compared to the \texttt{MINESweeper} metallicity. In general, the values show good agreement with the largest difference for the most metal-poor star being 0.51 dex. The final metallicities reported in this work is the average of these two values. }
    \label{fig:metcomp}
\end{figure}

In addition to the \texttt{MINESweeper} analysis, we further constrain the metallicities of these stars by modeling the Ca~II infrared triplet. This analysis provides an estimate of the [Ca/H] abundance which we then convert to [Fe/H], assuming a [Ca/Fe] abundance based on literature LMC abundances of similar metallicity stars \citep{Chiti2024}. 

In Figure \ref{fig:catrip}, we show the best-fit synthetic spectra model (blue dashed line) compared to the data (black line). The linearly interpolated synthetic spectra are described in Section \ref{sec:method}. We fit a continuum normalization for the model spectra to best match the observed spectra which is shown as a red dashed line. Although the entire region shown is used for fitting the continuum, we use only the two strongest Ca lines (8542 \AA\ and 8662 \AA) to fit the abundance. 

The comparison between the derived Ca triplet and \texttt{MINESweeper} metallicities are shown in Figure \ref{fig:metcomp}. They show good agreement with all differences within $\approx 2\sigma$. The star which shows the largest discrepancy (0.51 dex) is also the most metal-poor, and has a low SNR spectrum with few discernible absorption lines in the wavelength range fit by \texttt{MINESweeper}. For the final reported metallicity estimates we take the average of these two independent measurements and propagate the uncertainty accordingly.

\section{Fits to Carbon Molecular Features} \label{sec:app2}
\begin{figure*}
    \centering
    \includegraphics[width=\linewidth]{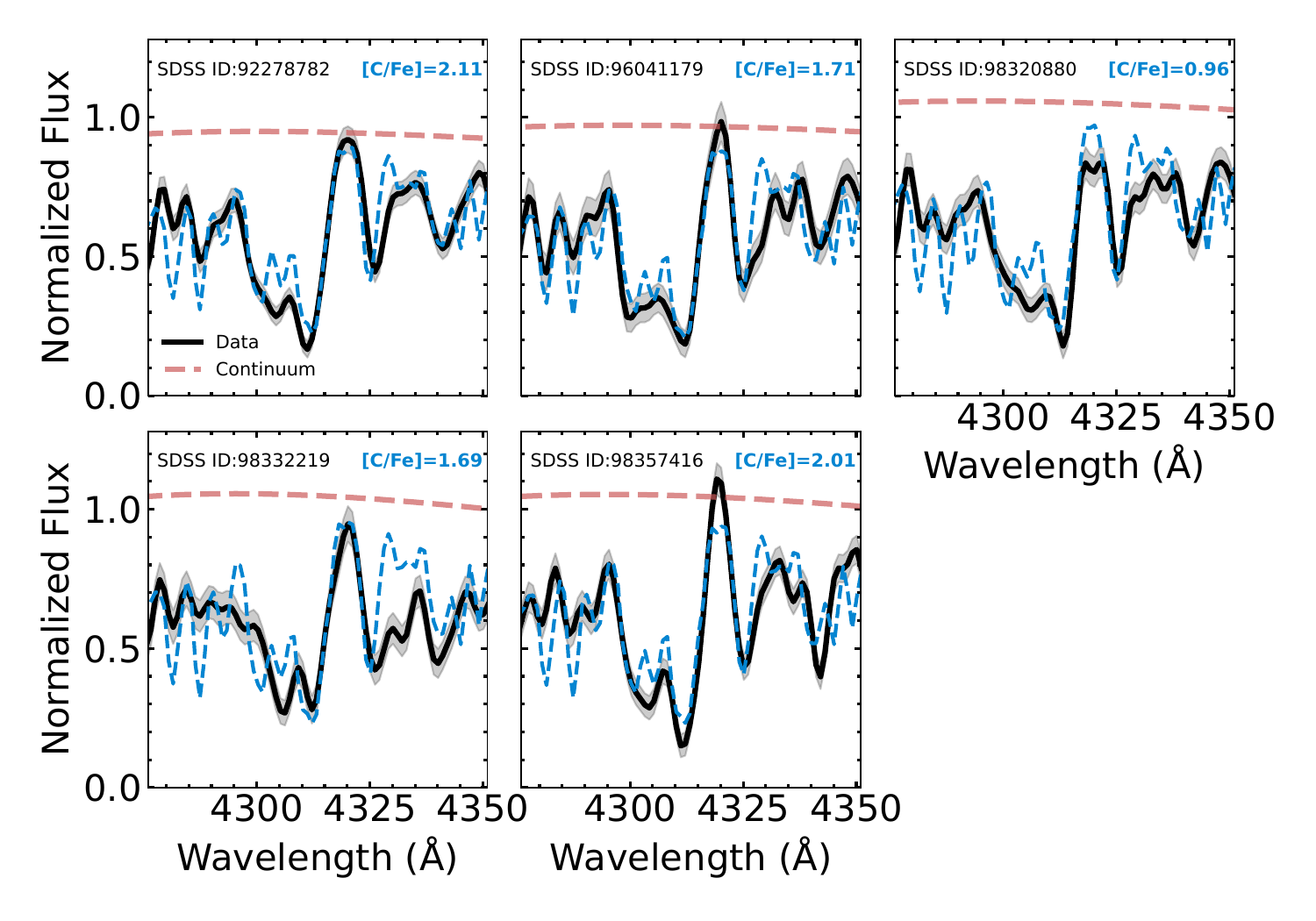}
    \caption{The fits to the CH G-band for the five CEMP stars. The black line with the grey shaded region give the observed normalized flux and the corresponding uncertainty. The blue dashed lines are the best fit synthetic spectra. As the fit is highly sensitive to the continuum normalization, we fit the model continuum to best match the observed spectra. The red dashed line gives the continuum normalization for the best-fit carbon abundance model.  }
    \label{fig:ch}
\end{figure*}

\begin{figure*}
    \centering
    \includegraphics[width=\linewidth]{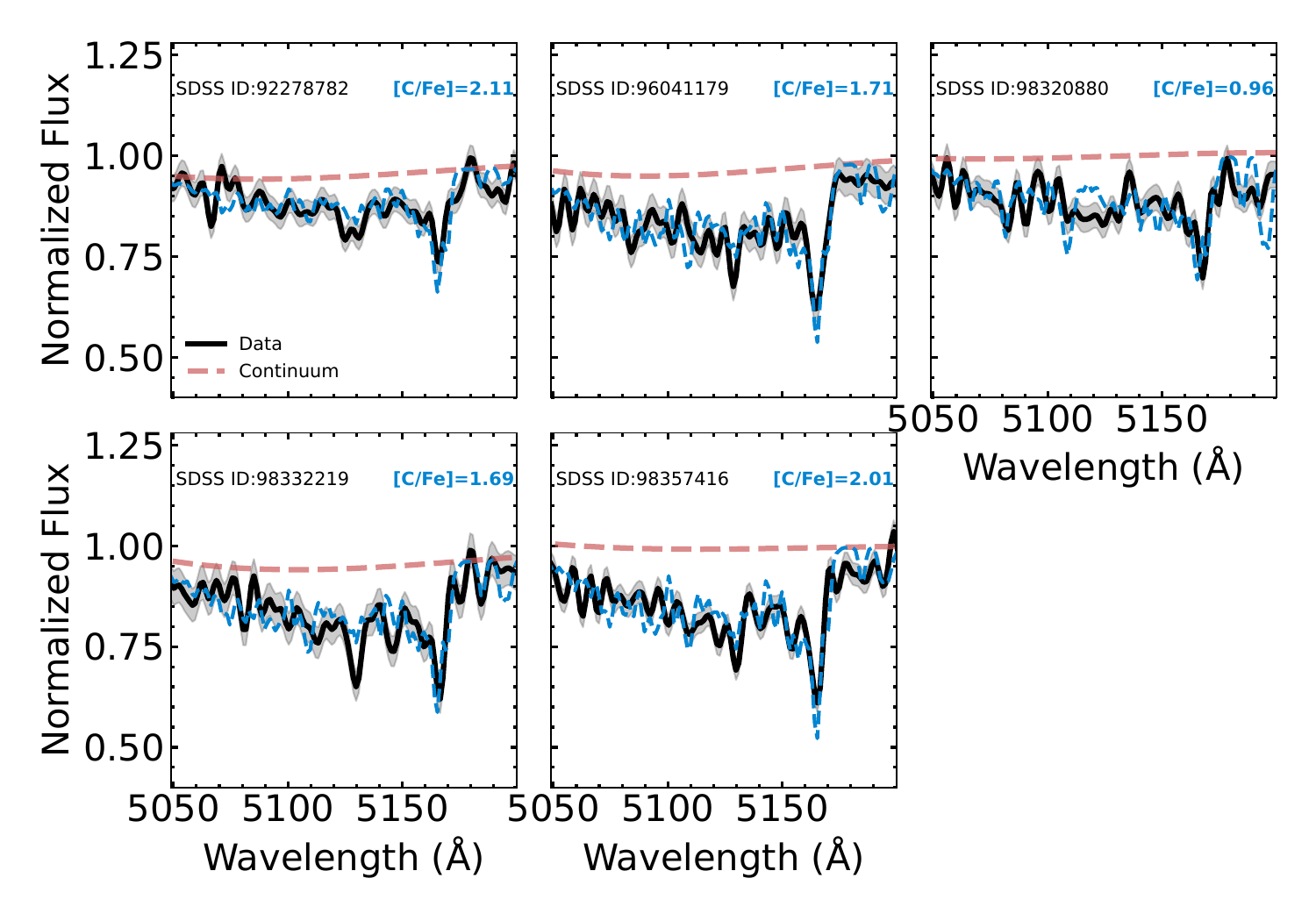}
    \caption{Same as Figure \ref{fig:ch}, but for the strongest $\rm{C_2}$ Swan band.   }
    \label{fig:c2}
\end{figure*}

To estimate the carbon abundance, we fit two molecular features, the CH G band at 4306 \AA\ and a $\rm{C_2}$ Swan band at 5165 \AA. In Figure \ref{fig:ch}, we show the fits to the CH G band while in Figure \ref{fig:c2} we show the $\rm{C_2}$ Swan band fits. In both figures, the observed spectra are shown as black lines and the best fit model spectra are shown as blue dashed lines. The fits are very sensitive to the continuum normalization. To account for this, we refit the continuum normalization for the model spectra (red dashed line) to best match the observed. 

\section{LMC Association}

\begin{figure}
    \centering
    \includegraphics[width=\linewidth]{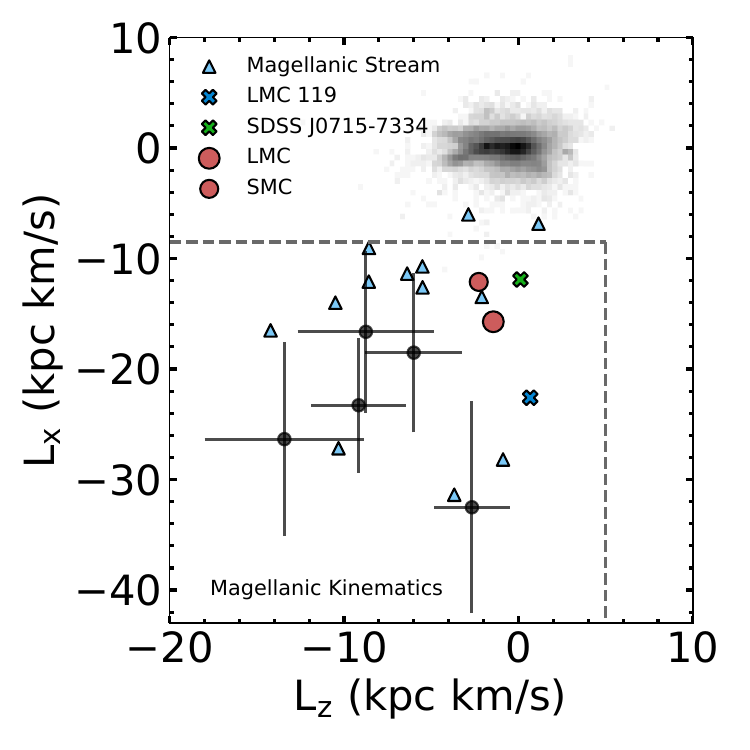}
    \caption{The Galactocentric specific angular momentum in the $\rm{L_x-L_z}$  plane for our five CEMP stars (black circles) compared to other Magellanic system objects and the SDSS-V DR20 MINESweeper halo catalog (background grey density). The LMC (large red circle) and SMC (smaller red circle) have large negative $\rm{L_x}$ relative to the Milky Way halo, making this a prime kinematic quantity for isolating stars associated with the Magellanic system. The Magellanic Stream stars (light blue triangles) and the Magellanic kinematic selection box defined by $\rm{L_x}<-8.5~\rm{kpc~km/s}$ and  $\rm{L_z}<5~\rm{kpc~km/s}$ are taken from \citet{Chandra2023}.    We also show recently discovered ultra-metal-poor stars associated with the LMC: LMC 119 (blue cross) and SDSS J-715-7334 (green cross).  }
    \label{fig:angmom}
\end{figure}

 To further verify that our five CEMP stars belong to the the LMC, we investigate their specific angular momentum in Galactocentric coordinates, following \citet{Chandra2023} and \citet{Ji2025}. The large negative $\rm{L_x}$ of the LMC and SMC makes this quantity especially useful for distinguishing between Magellanic system stars from Milky Way halo stars, which typically have ($\rm{L_x}$, $\rm{L_y}$, $\rm{L_z})\approx(0,0,0)~\rm{kpc~km/s}$. \citet{Chandra2023} therefore defined a kinematic selection for Magellanic system membership as $\rm{L_x}<-8.5~\rm{kpc~km/s}$ and  $\rm{L_z}<5~\rm{kpc~km/s}$,, which we show in Figure \ref{fig:angmom}. The background greyscale shows the density of SDSS-V DR20 MINESweeper halo stars in the $\rm{L_x}-\rm{L_v}$ plane. We also plot the specific angular momenta of LMC (large red circle) and SMC (smaller red circle), stars associated with the Magellanic stream from \citet{Chandra2023}, and the ultra-metal-poor LMC stars LMC 119 \citep[][blue X]{Chiti2024} and SDSS J0715-7334 \citep[][green X]{Ji2025}. All five of our CEMP stars (black circles) are consistent with LMC system membership.  Moreover, their angular momenta are more negative than the LMC, which provides especially strong evidence for association with LMC system. This is because angular momentum decays as the LMC falls into the Milky Way potential, and contamination from the Milky Way decreases with larger absolute values of $\rm{L_x}$, $\rm{L_y}$, and $\rm{L_z}$.

\bibliography{bibliography}{}
\bibliographystyle{aasjournal}

\end{document}